\documentclass{aa}
\usepackage{graphicx}
\usepackage{lscape}
\usepackage{longtable}
\usepackage{txfonts}
\usepackage{natbib}

\begin{document}

%\title{The XMM-Newton high-resolution X-ray spectrum of the classical T Tauri star BP Tauri:
%Evidence for an accretion shock}
%\title{Classical T Tauri stars: The Brothers of Accreting Neutron Stars}
\title{X-rays from accretion shocks in T Tauri stars: The case of BP Tau}
\author{J.H.M.M. Schmitt\inst{1}, J. Robrade\inst{1}, J.-U. Ness\inst{2}, 
F. Favata\inst{3}, B. Stelzer\inst{4}}
\institute{
% INST 1
Hamburger Sternwarte,
Gojenbergsweg 112,
D-21029 Hamburg, Germany
\and
% INST 2
University of Oxford, Department of Theoretical Physics,
1 Keble Road, Oxford OX1\,3NP, United Kingdom
\and
% INST 3
Astrophysics Division,
ESA/ESTEC,
P.O. Box 299, 2200 AG, Noordwijk,
The Netherlands
\and
% INST 4
Dipartimento di Scienze Fisiche e Astronomiche,
Universit\`a di Palermo, Piazza del Parlamento 1, I-90134 Palermo,
Italy
}

%\offprints{J. Schmitt}
%\mail{J. Schmitt, jschmitt@hs.uni-hamburg.de}
\titlerunning{X-ray emission from an accretion shock on BP Tau}
\date{Received / Accepted }

\abstract{We present an XMM-Newton observation of the classical T Tauri star BP Tau. In the XMM-Newton
RGS spectrum the O\,{\sc vii} triplet is clearly detected with a very weak forbidden line 
indicating high plasma densities and/or a high UV flux environment. 
At the same time concurrent UV data point to a
small hot spot filling factor suggesting an accretion funnel shock as the site of the X-ray and UV
emission.  Together with the X-ray data on TW Hya these new observations suggest such funnels to be a general 
feature in classical T~Tauri stars.
\keywords{X-rays: stars -- stars: individual: BP\,Tau -- stars: pre-main sequence, coronae,
activity -- accretion}}

\maketitle

\section{Introduction}

One of the central results of stellar X-ray astronomy is the discovery that
all ``cool stars'', i.e., stars with outer convective envelopes, are surrounded 
by hot coronae \citep{schmitt04}.
%"cool stars" are stars with outer convective envelopes, i.e.,
%stars with spectral type $\approx$ A7 down to the very bottom of the main sequence.
The usual interpretation of this finding is that the combined action of turbulence in the outer
convection zones and the ever-present rotation leads to dynamo action
with vigorous production of magnetic fields and ensuing activity. X-ray observations
support this interpretation with the ``onset of activity'', i.e.,
the rapid increase in X-ray detection rates of main sequence stars at spectral type
$\approx$ F0 and the correlation between rotation rate and activity.

Strong X-ray emission is found also from many young stars such as zero age main sequence stars
or T~Tauri above the main sequence. Both flavors of T~Tauri stars, i.e., the weak line
T~Tauri stars (wTTS) without disks as well as the classical T Tauri stars (cTTS) with disks 
are X-ray sources
\citep{feig99}. The presence of a disk around cTTS constitutes a fundamental difference compared to
a ``normal'' star without disk, since the energy stored in the disk material can be released
through accretion and lead to super-photospheric and/or time variable emissions; in fact, the
optical variability observed in (some) cTTS \citep{gull96} is commonly interpreted
in this fashion. Yet the high-energy activity observed from cTTS and wTTS is usually
interpreted as a ``scaled-up'' version of solar activity. The low-resolution X-ray data available
prior to XMM-Newton and {\it Chandra} made a distinction between the X-ray properties of wTTS and cTTS
very difficult; while some differences in emission level and variability properties
between wTTS and cTTS were indicated by statistical studies
\citep{stelzer00,stelzer01,flac2003}, a characterization of the physical properties of the
X-ray emitting regions was impossible.

This situation has changed with the advent of high-resolution grating spectroscopy with
XMM-Newton and {\it Chandra}. A great surprise was
the X-ray spectrum of the cTTS TW Hya \citep{kastner02,stelzer04}, which clearly showed
low forbidden to intercombination (f/i) line ratios in the O\,{\sc vii} and Ne\,{\sc ix} triplets,
which were interpreted as due to high plasma densities in the X-ray emitting regions.
The extensive spectral survey by \cite{ness04}, who studied high-resolution X-ray spectra 
of 48 coronal sources, provided no star that would even come close
to the low O\,{\sc vii} f/i-ratio as observed for TW Hya. 
Clearly, TW Hya differs in its coronal properties
from all other stars, and it is very natural to ascribe this difference to the presence of an
accretion disk around TW~Hya. Specifically, \cite{kastner02} and \cite{stelzer04} interpret the
X-ray emission as arising from an accretion shock produced by matter falling down onto TW~Hya along
a magnetic funnel. There is an obvious need to examine further cTTS  with high-resolution
spectroscopy in order to assess whether TW Hya's X-ray properties are typical for cTTS or not.
%regarding its X-ray properties or not. 
We therefore obtained a high-resolution X-ray spectrum of the cTTS BP Tau
with the Reflection Grating Spectrometer (RGS) onboard XMM-Newton, and the purpose of this letter
is to present and discuss the implications of this observation.
\vspace{-0.5cm}
\section{Observations}

\subsection{The cTTS BP Tau: Optical and X-ray properties}

An extensive overview of the optical properties of BP Tau is given by \cite{gull96} and \cite{errico01}.
Its spectral type varies from K5-K7, it rotates fast (v sin(i) = 15.4 km/sec), but various
rotation period measurements varying between 6.1 to 8.3 days have been published.
The cTTS nature of BP Tau is clearly demonstrated by the excess continuum (veiling) and Balmer
emission \citep{bertout88}. There is debate about BP Tau's distance, and \cite{wichmann98} argue
against BP Tau being an outlier from the Taurus-Auriga cloud as suggested by its HIPPARCOS parallax.
From extensive optical monitoring \cite{gull96} conclude
that optical variability is common in BP Tau, but in character very much different from variability
encountered in typical flare stars. Most of the observed changes are slow and smooth and are
interpreted as the result of inhomogeneous accretion from the disk onto the stellar surface.
This view is strongly supported by the circular polarization in the He\,{\sc i}
$\lambda$ 5876 emission line measured by \cite{john99}, who 
deduce a mean longitudinal magnetic field of $2460\,\pm\,120$\,G
in the line forming region and argue that
accretion occurs preferentially along large-scale magnetic loops with a small filling factor.
X-ray emission from BP Tau at a level of $\approx 10^{30}$\,erg/sec
was first reported by \cite{walter81} using the {\it Einstein Observatory}.
In a simultaneous optical and X-ray observations of BP Tau with ROSAT \cite{gull97} 
found no evidence for correlated variations between the two bands and thus 
attributed the X-ray emission to magnetically active regions.

\begin{center}
\begin{figure}[h]
\includegraphics[scale=0.5]{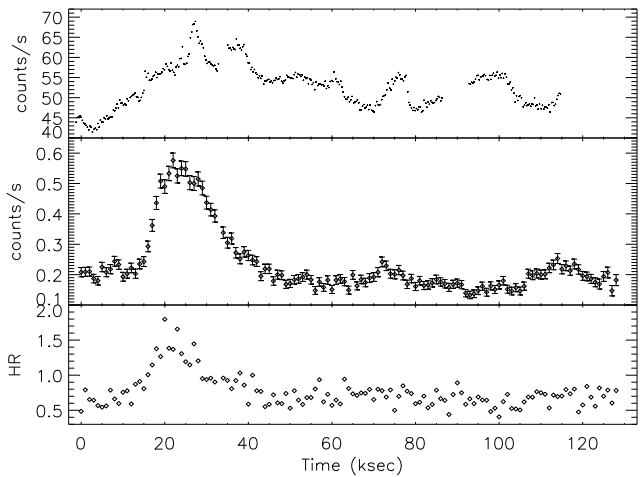}
\caption{
OM UVW1 light curve (upper panel; bin size 300 sec),
EPIC PN X-ray light curve (medium panel), and X-ray hardness ratio defined as
ratio between the rates in the bands 0.2-1.0 and 1.0-10.0 keV
(lower panel; both with bin size 1000 sec). \label{fig1}}
\end{figure}
\end{center}
\vspace{-1.5cm}
\subsection{XMM-Newton X-Ray Data}

BP Tau was observed with XMM-Newton on August 15, 2004 for a duration of
131\,ksec (Obs-ID 0200370101) with the RGS as prime instrument.
The observations were performed in full-frame mode employing the thick filter for both the MOS
and the PN cameras of the European Photon Imaging Camera (EPIC). 
The optical monitor (OM) was operated using the UVW1 filter with a
band pass between 2500\,\AA\ and 3500\,\AA\ and an effective wavelength of 2910\,\AA \  
according to the XMM-Newton users' handbook (\cite{xmm}). All XMM-Newton data was analyzed with 
SAS version 6.0.
Background conditions were very quiet throughout most of the BP Tau observation and only
10\,ksec of data had to be screened.
In Fig.~\ref{fig1} we show the EPIC PN X-ray light curve, the X-ray hardness ratio (lower panel),
and the UVW1 light curve (upper panel). An obvious X-ray flare with a spectral hardness increase occurred 
between $\approx 20-40$\,ksec into the observations; the X-ray flare may be accompanied by a much longer lasting
UV event, but there is no strict correlation between X-ray and UV variability. 
\begin{center}
\begin{figure}[h]
\includegraphics[scale=0.35,angle=-90.]{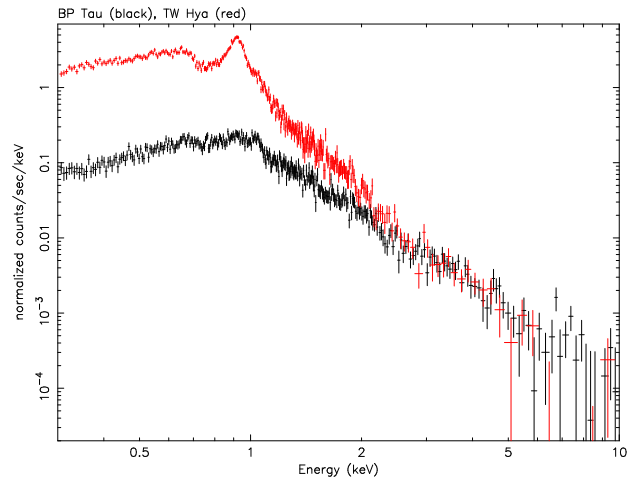}
\caption{EPIC PN spectrum of BP Tau (quiescent phase; lower black points) in comparison to the EPIC PN 
spectrum of TW Hya (upper, red/grey points) \label{fig2}}
\end{figure}
\end{center}
\vspace{-1.cm}
In Fig.~\ref{fig2} we show the EPIC PN spectrum of BP Tau (lower curve) in comparison to TW Hya
\citep[upper curve; cf. ][]{stelzer04}. Above $\approx 2.5$\,keV the two spectra
overlap, while at lower energies the flux of TW Hya exceeds that of BP Tau by almost an order of magnitude.
Spectral modeling of the EPIC PN spectrum using simple multi-temperature fits
requires absorption column densities of $1-2\times 10^{21}$\,cm$^{-2}$ depending on the chosen model and
consistent with the relation between $N_{\rm H}$ and $E_{\rm B-V}$ \citep{jenkins74};
hot temperatures ($\approx 2$\,keV) are required in
contrast to TW Hya, whose EPIC spectrum is dominated by cool plasma of
$\approx 0.3$\,keV \citep[cf.,][]{stelzer04}. Thus the EPIC spectra of BP Tau and TW Hya
suggest differences between the two stars.  
\begin{center}
\begin{figure}[h]
\includegraphics[scale=0.49]{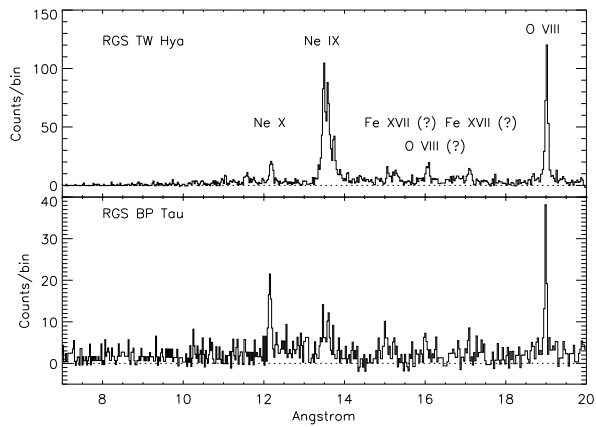}
\caption{Comparison of instrumental background subtracted RGS2 spectra of TW Hya and BP Tau with
line identifications. \label{fig5}}
\end{figure}
\end{center}
\vspace{-0.5cm}
However, BP Tau's instrumental background-subtracted high resolution RGS2 spectrum looks 
rather similar to that of TW Hya, (cf., Fig. \ref{fig5}); probing the lower-energy plasma
it shows mostly oxygen and neon lines with -- at best -- weak
iron lines as well as signs of a continuum.  
Because of the low signal only a few spectral lines can be reliably
detected in our RGS spectra. The clearly detected lines include the O\,{\sc viii}
Ly$_\alpha$ line, the O\,{\sc vii} triplet, the Ne\,{\sc x}
Ly$_\alpha$ line, and the Ne\,{\sc ix} triplet at 13.5\,\AA;
interestingly, no unambiguous RGS detections of iron lines, in particular at 15.03\,\AA\
and 17.07\,\AA\ were obtained (cf., Fig. \ref{fig5} and Table \ref{tab1}), 
reminiscent of the low amount of iron 
derived for TW Hya. Also, no lines of nitrogen and carbon could be detected,
presumably because of the larger absorption column towards BP Tau.
The detected lines together with the derived best fit line counts and
their errors are listed in Table~\ref{tab1}; all line counts were
derived with the CORA program \citep{ness02}, assuming Lorentzian line shapes.
The line fits were carried out in such a way that the wavelength differences between the triplet lines
and line widths were held constant; the background was adjusted by eye, different
choices of background lead to line flux variations well within the errors; in 
Fig.~\ref{fig3} we plot the RGS spectrum around the O\,{\sc vii} triplet at 22\,\AA \ together 
with our best fit model.
% Fig. \ref{fig4} shows the same information for
%the Ne\,{\sc ix} triplet. For both triplets we find clear detections of the resonance and intercombination
%lines, but only marginal detections of the forbidden line (see also Table \ref{tab1}) .
From the numbers listed in Table~\ref{tab1} we find an observed f/i-ratio of $0.37\,\pm\,0.16$ for O\,{\sc vii} 
and $0.40\,\pm\,0.26$ for Ne\,{\sc ix}; the latter assumes negligible contamination 
by iron, which is a severe problem for the interpretation of any Ne\,{\sc ix} triplet data \citep[cf., ][]{ness03}
depending on the strength of the iron lines.

\begin{table} [h]
\begin{tabular}{| l l r r r r|}
\hline
Line ID & $\lambda$ (\AA)  & Counts & Error & Instr. & Flux \cr
\hline
Ne\,{\sc x} Ly$_\alpha$ & 12.14  & 91.5 & 11.7 & RGS2 & 13.2 \cr
Ne\,{\sc ix} He r & 13.46 & 37.0 & 9.9 & RGS2 & 5.2 \cr
Ne\,{\sc ix} He i & 13.56 & 31.2 & 9.8 & RGS2 & 4.4 \cr
Ne\,{\sc ix} He f & 13.71 & 12.5 & 7.1 & RGS2 & 1.7 \cr
O\,{\sc viii} Ly$_\beta$ & 16.03 & 25.8 & 6.4 & RGS1 & 4.1 \cr
O\,{\sc viii} Ly$_\beta$ & 16.00 & 16.7 & 6.2 & RGS2 & 2.3 \cr
O\,{\sc viii} Ly$_\alpha$ & 18.97 & 95.5 & 11.4 & RGS2 & 14.2 \cr
O\,{\sc vii} He r & 21.6 & 47.7 & 8.6 & RGS1 & 8.9 \cr
O\,{\sc vii} He i & 21.8 & 36.6 & 7.8 & RGS1 & 7.1 \cr
O\,{\sc vii} He f & 22.1 & 13.7 & 5.5 & RGS1 & 2.7 \cr
Fe\,{\sc xvii}  & 15.03 & $<$ 20&  & RGS1 & $<$ 2.9\cr
\hline
\end{tabular}
\caption{X-ray lines detected in the RGS spectrum of BP Tau, line counts
and photon fluxes in units of $10^{-6}$\,ph\,s$^{-1}$\,cm$^{-2}$ \label{tab1}}
\end{table}
\vspace{-1cm}
\begin{center}
\begin{figure}[h]
\includegraphics[scale=0.49]{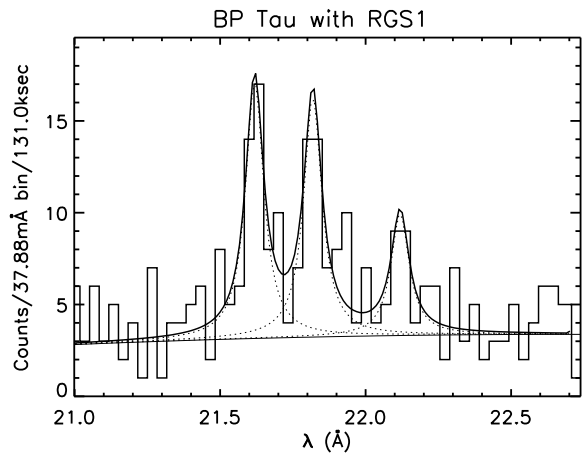}
\caption{RGS spectrum BP Tau: O\,{\sc vii} triplet region with best fit model.\label{fig3}}
\end{figure}
\end{center}
\vspace{-1.0cm}
%\section{Discussion and Conclusions}
\section{Discussion}
\cite{stelzer04} interpret the XMM-Newton X-ray data on the cTTS TW Hya in terms of an accretion funnel scenario,
where the X-ray emission is emitted in a shock (``hot spot'')
produced by the infall of material  along the magnetic field at essentially free-fall
velocity onto the stellar surface. Does the same scenario also apply to the cTTS BP Tau
and possibly to cTTS as a class ?

It is instructive to compare the observed line fluxes from BP Tau with those measured from TW Hya
\citep[cf., Table~2 in ][]{stelzer04}, taking into account the different amounts of interstellar
absorption.  For ISM columns of $3\times 10^{20}$\,cm$^{-2}$ and $1\times 10^{21}$\,cm$^{-2}$
respectively, we compute ISM transmissivities of (0.83, 0.54) at 18.97\,\AA, (0.77, 0.42) at 21.6\,\AA,
(0.93, 0.78) at 12.14\,\AA, and (0.91, 0.73) at 13.46\,\AA\ for TW Hya and BP Tau respectively. These values lead to
ratios between the Ly$_{\alpha}$ and He-like r flux ratios 1.9 and 2.1 for oxygen and
0.3 and 2.7 for neon respectively. While the ratios for oxygen are about similar, those for neon
differ quite significantly, thus the emission measure distribution in BP Tau
must also be different. We also note that the Ly$_{\alpha}$ and He-like r flux ratios observed in BP Tau 
for oxygen and neon are not consistent with a single temperature.  

\begin{table} [h]
\begin{tabular}{| r r r r |}
\hline
Line & low state & high state & ``low state''/3.2 \cr
\hline
Ne\,{\sc x} Ly$_\alpha$   & 48.6 $\pm$ 8.7& 42.7 $\pm$ 8.0&  15.2  $\pm$ 2.7\cr
O\,{\sc viii} Ly$_\alpha$ & 59.2 $\pm$ 9.1& 32.5 $\pm$ 6.6&  18.5  $\pm$ 2.8\cr
O\,{\sc vii} He r         & 34.4 $\pm$ 7.3& 10.7 $\pm$ 4.1&  10.8  $\pm$ 2.3\cr
O\,{\sc vii} He i         & 24.6 $\pm$ 6.3&  9.5 $\pm$ 4.2&  7.7   $\pm$ 2.0\cr
O\,{\sc vii} He f         & 5.5  $\pm$ 3.8&  5.6 $\pm$ 3.2&  1.7   $\pm$ 1.2\cr
\hline
\end{tabular} 
\caption{Line counts measured in ``low state'' and in  ``high state'' and extrapolated
``high state'' counts.\label{tab2}}
\end{table}
\vspace{-0.5cm}
We next checked to what extent individual RGS-detected lines are influenced by
the flare.  We defined a ``high state'' by including all data recorded
in the time interval 13.5 - 43.0 ksec (counted from the start of the observations as in Fig. \ref{fig1}) 
and a ``low state'' including all the rest and determined line
counts as before.  We also computed the expected number of ``high state'' counts extrapolating
from the ``low state'' values and list all results in Table \ref{tab2}.   Clearly,
within the errors the  O\,{\sc vii} line counts are unaffected by the flaring emission,
and only 15 \% of the recorded O\,{\sc viii} line counts can be attributed to the flare; for
Ne\,{\sc x} this contribution rises to 30 \%.  We thus conclude that the observed O\,{\sc vii} emissions
are not significantly affected by the flare, while O\,{\sc viii} Ly$_\alpha$ is contaminated by $<$ 20 \%.
%Given the low statistics
%it is impossible to disentangle possible contributions to the oxygen and neon lines from hotter (flaring) 
%plasma in the RGS spectrum.  Modeling of the EPIC data
%shows the largest emission measure components residing at low temperatures, suggesting that the
%oxygen emission predominantly comes from the coolest plasma components.

In addition to the X-ray data concurrent photometric UV data from the XMM-Newton OM are available
for BP Tau (cf., Fig.~\ref{fig1}). The UV continuum flux of BP Tau is thought to be
produced by an accretion hot spot as demonstrated by \cite{ardila00} and \cite{errico01}.
The same applies of course also to our OM UVW1 photometric data, which would require filling factors
of ten or more stellar surfaces if they were to be reconciled with BP Tau's photospheric temperature. 
We therefore assume in the following that the recorded UVW1 flux is exclusively produced
in a hot spot and deredden this UV-flux using $A_{\rm V} = 0.51$ \citep{gull98} and $E_{\rm B-V} = 0.32 \times A_{\rm V}$.
Without any further spectral information we assume an emergent blackbody spectrum with variable
temperature $T_{\rm em}$, convert count rate into specific energy flux (erg/cm$^2$/s/\AA)
using the instrumental conversion factor of $4.4\times 10^{-16}$ (erg/cm$^2$/count/\AA), and compute
the fractional surface area and filling factor required to account for the (dereddened) observed UVW1 flux.
Assuming hot spot temperatures between 8000 - 10000\,K \citep[cf., ][]{calvet98,ardila00},
we invariably find filling factors between
0.6 -- 4\%, which are in line with values computed from more sophisticated models including shocks
and an irradiated photosphere \citep{calvet98}. 

Converting the observed oxygen f/i-ratio of $0.37\,\pm\,0.16$ to density yields a nominal value
of log $n_e = 11.48$ with a (formal) error range of (11.3-11.8), neglecting radiative
deexcitation of the forbidden line level; therefore the derived densities  may be regarded as
upper limits if the X-ray emitting region is affected by the observed UV radiation.
%Applying the same procedure to neon, results in a density of log $n_e = 12.65$ with an error 
%range of (12.4--13.1), and 
We re-emphasize that both BP Tau and TW Hya have lower f/i-ratios and higher densities than all of the
coronal X-ray sources analyzed by \cite{ness04}.
If the O\,{\sc vii} f/i-ratio was contaminated by a lower densisty ``coronal component'', 
any accretion-related shock component would require an even smaller f/ir-ratio.

Assuming optically thin emission, the observed energy flux in the O\,{\sc vii} {\bf and} 
O\,{\sc viii} lines, $f_{\rm oxy}$, is given by 
$f_{\rm oxy} = A_{\rm sh} z n_{\rm e}^2 P_{\rm oxy}/4 \pi d^2$, where $A_{\rm sh}$, $z$, $n_{\rm e}$, $ P_{\rm oxy}$ denote shock area, cooling zone
thickness, electron density and line cooling function respectively.  The post shock plasma cooling time $\tau$ 
is given by $\tau = 3 n_e k T/n_e^2 P_{\rm tot}$, with $k$, $T$ and $P_{\rm tot}$ denoting  Boltzmann's constant, temperature and
overall radiative loss function.  Postshock velocity $V_{\rm post}$,  $z$ and  $\tau$ are related through $\tau = z/V_{\rm post}$.
Thus we calculate the mass accretion rate 
$M_{\rm acc} = \frac {4 \pi d^2 f_{\rm oxy} m_H \mu  P_{\rm tot}} {3 k T P_{\rm oxy}}$, with $m_{\rm H}$ denoting 
hydrogen mass  and $\mu$ the mean molecular weight.  With the observed values for $f_{\rm oxy}$ we find with T = 2.5 MK
$M_{\rm acc} \approx 9 \times 10^{-10}$ M$_{\odot}$/yr, about an order of magnitude smaller than inferred at UV wavelengths;
since the observed oxygen fluxes may contain non-accretion related contributions, this value of
$M_{acc}$ should be considered as an upper limit.
Filling factors of a few percent yield densities consistent with the O\,{\sc vii} triplet and a cooling
zone thickness of $\approx$ 100 km.  As to the depth of the X-ray emitting region,
a model independent absorption estimate using the observed
photon flux ratio in the Ly$_\alpha$ and Ly$_\beta$ lines of 3.1 (cf., Table~\ref{tab1})
requires with temperature of $\log T = 6.8$ an
equivalent absorption column of $\approx 5\times 10^{21}$\,cm$^{-2}$. Given the large errors
on the Ly$_\beta$/Ly$_\alpha$ photon flux ratio and the internal consistency between RGS2 values  
and optical extinction and the $N_{\rm H}$ values of
$N_H \approx 2 \times $10$^{21}$cm$^{-2}$ derived from the broad band X-ray spectra, we 
conclude that there is
no real evidence for an ``additional'' absorption of BP Tau's X-ray flux.
%\vspace{-0.5cm}
\section{Conclusions}
The XMM-Newton RGS spectrum clearly demonstrates that X-ray emitting O\,{\sc vii} layers in the cTTS BP Tau
are either at high density and/or immersed in a strong ultraviolet flux. Both possibilities can be well
explained by an X-ray emitting accretion shock on BP Tau.  This accretion shock can, however, produce 
only the low-temperature components in the broad band X-ray spectra of BP Tau (and TW Hya); to explain the 
high temperature component additional processes possibly involving magnetic interactions between the disk 
and the star or magnetic activity anchored in the photosphere are required.  Also, the derived mass 
accretion rates are smaller than those inferred from
optical and UV data, but a detailed modeling in particular taking into account non-equilibrium effects
is still lacking.  At any rate, the XMM-Newton observations of BP Tau show that 
TW Hya is not ``alone''. Accretion shocks at the end of magnetic funnels
connecting disk and stellar surface may in fact be a common feature of cTTS stars as a class.
\vspace{-0.15cm}
\begin{acknowledgements}

This work is based on observations obtained with {\em XMM}-Newton,
an ESA science mission with instruments and contributions directly
funded by ESA Member States and the USA (NASA).
JR and JUN acknowledge support from DLR under grant 50OR0105 and 
PPARC under grant PPA/G/S/2003/00091.
\end{acknowledgements}
%\vspace{-0.75cm}


\begin{thebibliography}{}

\bibitem[\protect\astroncite{Ardila \& Basri}{2000}]{ardila00}
Ardila, D., R., Basri, G.S. 2000, ApJ, 539, 834

\bibitem[\protect\astroncite{Bertout et al.}{1988}]{bertout88}
Bertout, C., Basri, G.S., Bouvier, J. 1998, ApJ, 330, 350

\bibitem[\protect\astroncite{Calvet \& Gullbring}{1998}]{calvet98}
Calvet, N. \& Gullbring, E. 1998, ApJ, 509, 802

\bibitem[\protect\astroncite{Ehle et al.}{2004}]{xmm}
Ehle, M., Breitfellner, M., Gonzales Riestra, M., et al. 2004, XMM-Newton User's Handbook

\bibitem[\protect\astroncite{Errico et al.}{2001}]{errico01}
Errico, L., Lamzin, S.~A., {Vittone}, A.~A., 2001, A\&A, 377, 557

\bibitem[\protect\astroncite{Gullbring et al.}{1996}]{gull96}
Gullbring, E., Barwig, H., Chen, P.S., Gahm, G.F., Bao, M.X. 1996,
A\&A, 307, 791

\bibitem[\protect\astroncite{Gullbring et al.}{1997}]{gull97}
Gullbring, E., Barwig, H., Schmitt, J.H.M.M. 1997, A\&A, 324, 155

\bibitem[\protect\astroncite{Gullbring et al.}{1998}]{gull98}
Gullbring, E., Hartmann, L., Brice\~no, C., Calvet, N. 1998, ApJ, 492, 392

\bibitem[\protect\astroncite{Feigelson \& Montmerle}{1999}]{feig99}
Feigelson, E.D. \& Montmerle, T. 1999, Ann.Rev.A\&A, 37, 363

\bibitem[\protect\astroncite{Flaccomio et al.}{2003}]{flac2003}
Flaccomio, E., Micela, G., Sciortino, S. 2003, A\&A, 402, 277

\bibitem[\protect\astroncite{Jenkins \& Savage}{1974}]{jenkins74}
Jenkins, E.B., Savage, B.D. 1974, ApJ, 187, 243

\bibitem[\protect\astroncite{Johns-Krull et al.}{1999}]{john99}
Johns-Krull, C.M., Valenti, J.A., Hatzes, A.P., \& Kanaan, A. 1999, ApJ, 510, L41

\bibitem[\protect\astroncite{Kastner et~al.}{2002}]{kastner02}
Kastner, J.H., Huenemoerder, D.P., Schulz, N.S., \& Canizares, C.R. 2002, ApJ, 567, 434

\bibitem[\protect\astroncite{Ness \& Wichmann}{2002}]{ness02}
Ness, J.-U. \& Wichmann, R. 2002, AN, 323, 129

\bibitem[\protect\astroncite{Ness et~al.}{2003}]{ness03}
Ness, J.-U., Brickhouse, N.S., Drake, J.J., \& Huenemoerder, D.P. 2003, ApJ, 598, 1277

\bibitem[\protect\astroncite{Ness et~al.}{2004}]{ness04} Ness, J.-U.,
G\"udel, M., Schmitt, J.H.M.M., et al. 2004, A\&A, 427, 667

\bibitem[\protect\astroncite{Schmitt \& Liefke}{2004}]{schmitt04}
Schmitt, J.H.M.M. \& Liefke, C. 2004, A\&A, 417, 651

\bibitem[\protect\astroncite{Stelzer et~al.}{2000}]{stelzer00}
Stelzer, B., Neuh\"auser, R. \& Hambaryan, V. 2000, A\&A, 356, 949

\bibitem[\protect\astroncite{Stelzer \& Neuh\"auser}{2001}]{stelzer01}
Stelzer, B. \& Neuh\"auser, R. 2001, A\&A, 377, 538

\bibitem[\protect\astroncite{Stelzer \& Schmitt}{2004}]{stelzer04}
Stelzer, B. \& Schmitt, J.H.M.M. 2004, A\&A, 418, 687

\bibitem[\protect\astroncite{Walter \& Kuhi}{1981}]{walter81}
Walter, F.M \& Kuhi, L.V. 1981, ApJ, 284, 194

\bibitem[\protect\astroncite{Wichmann et al.}{1998}]{wichmann98}
Wichmann, R., Bastian, U., Krautter, J., Jankovics, I., \& Rucinski, S.M. 1998, MNRAS, 301L, 39


\end{thebibliography}
\end{document}